\documentclass[
twocolumn,
prl,showpacs,preprintnumbers,amsmath,amssymb]{revtex4}

\usepackage{times}
\usepackage{graphicx}
\usepackage{amssymb}
\usepackage{amsthm}
\usepackage{amsmath}
\usepackage{dsfont}
\usepackage{bm}
\usepackage{mathrsfs}
\usepackage{bbold}

\begin{document}

\title{Casimir stress in materials: hard divergency at soft walls}
\author{Itay Griniasty and Ulf Leonhardt}
\affiliation{
\normalsize{
Department of Physics of Complex Systems,
Weizmann Institute of Science, Rehovot 761001, Israel}
}
\date{\today}

\begin{abstract}
The Casimir force between macroscopic bodies is well understood, but not the Casimir stress inside bodies. Suppose empty space or a uniform medium meets a soft wall where the refractive index is continuous but its derivative jumps. For this situation we predict a characteristic power law for the stress inside the soft wall and close to its edges. Our result shows that such edges are not tolerated in the aggregation of liquids at surfaces, regardless whether the liquid is attracted or repelled. 
\end{abstract}

\maketitle

\begin{figure}[t]
\begin{center}
\includegraphics[width=20.0pc]{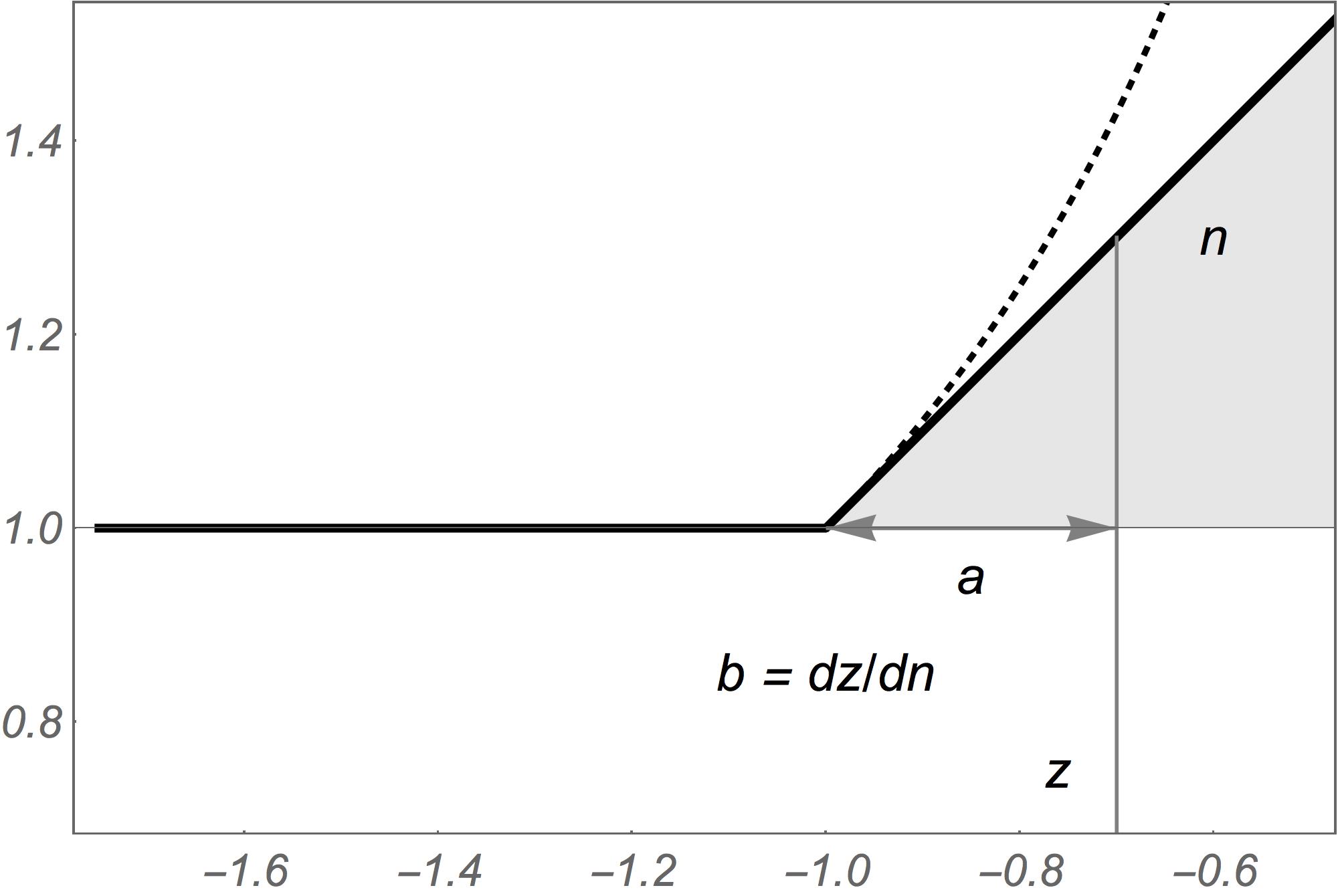}
\caption{
\small{
Soft wall. Refractive-index profile $n(z)$ of a planar material where the first derivative of $n$ jumps by $1/b$ at the edge to free space with $n=1$. Equation (\ref{result}) describes the Casimir stress near the edge with $a$ being the distance from the edge. The dotted line indicates the Beltrami profile of Eq.~(\ref{beltrami}), fitting the actual profile at the edge, employed to calculate the stress analytically. 
}
\label{softwall}}
\end{center}
\end{figure}

In 1948 Casimir \cite{Casimir} found an enigmatic formula for the part of the zero--point energy density of the electromagnetic field between two perfect mirrors that can do physical work:
\begin{equation}
U = \frac{\pi^2}{240}\,\frac{\hbar c}{a^4}
\label{casimir}
\end{equation}
where $a$ is the distance between the mirrors, $\hbar$ Planck's constant divided by $2\pi$, and $c$ the speed of light in vacuum. These days, nearly 70 years later, the field of Casimir forces is an established research area where modern theory \cite{Review,Reid} can predict the results of high--precision experiments with good accuracy. The Casimir force {\it between} macroscopic bodies is well understood, but surprisingly \cite{SimpsonSurprise}, not the force {\it inside} bodies. Only very recently, after several attempts \cite{Attempts,PXL} of establishing a theory of the Casimir stress inside materials, one was found \cite{Grin} that appears to be satisfactory. Here we report on the first prediction of that theory: the Casimir stress $\sigma$ near the edge of a soft wall \cite{Soft} (Fig.~\ref{softwall}) behaves like
\begin{equation}
\sigma_{zz} = \frac{23}{240\,(2\pi)^2}\,\frac{\hbar c}{a^2 b^2}
\label{result}
\end{equation}
where $\sigma_{zz}$ is the physically relevant stress component and $a$ denotes the distance from the edge to empty space. Here the refractive index $n$ changes continuously along the $z$ coordinate while its first derivative jumps by $1/b$ (Fig.~\ref{softwall}).

Equation (\ref{result}) shows that at discontinuities of the derivative of $n$, the Casimir stress diverges with a characteristic power law. In contrast, at discontinuities of $n$ itself, the physically relevant Casimir stress $\sigma_{zz}$ does not diverge, but merely jumps \cite{Remark}, and gives Eq.~(\ref{casimir}) for two plates with $n\rightarrow\infty$. Note that the divergency of the stress at the edge of the soft wall is a physical effect, not an artefact of the infinite bare zero--point energy that is removed in the renormalization of the Casimir force \cite{Grin}. The infinite physical stress at the edge implies that a discontinuity of the derivative of $n$ is not tolerated in liquids: if, for example, a liquid aggregates as a soft wall on a boundary, such discontinuities are immediately removed by the force density $\bm{\nabla}\cdot\sigma$ putting the liquid into motion. Discontinuities of $n$, on the other hand, are locally stable, leading only to forces between bodies and not to tension inside. Our result thus shows a striking feature in the aggregation process: regardless whether $n$ rises or falls, {\it i.e.} regardless whether the Casimir force is attractive or repulsive, a liquid cannot tolerate the edge of a soft wall; preferably it will form a discontinuity of the refractive index: either it will be aggregated or repelled. This application of the theory of Casimir forces inside materials \cite{Grin} resembles the early tests of the Lifshitz theory \cite{Lifshitz} of forces between materials in the wetting of surfaces \cite{DLP}. There the Casimir stress at the interface between a liquid and a solid wall gives the wetting angles of droplets on the surface, here the Casimir stress inside the liquid describes the consolidation of surfaces.

\begin{figure}[t]
\begin{center}
\includegraphics[width=20.0pc]{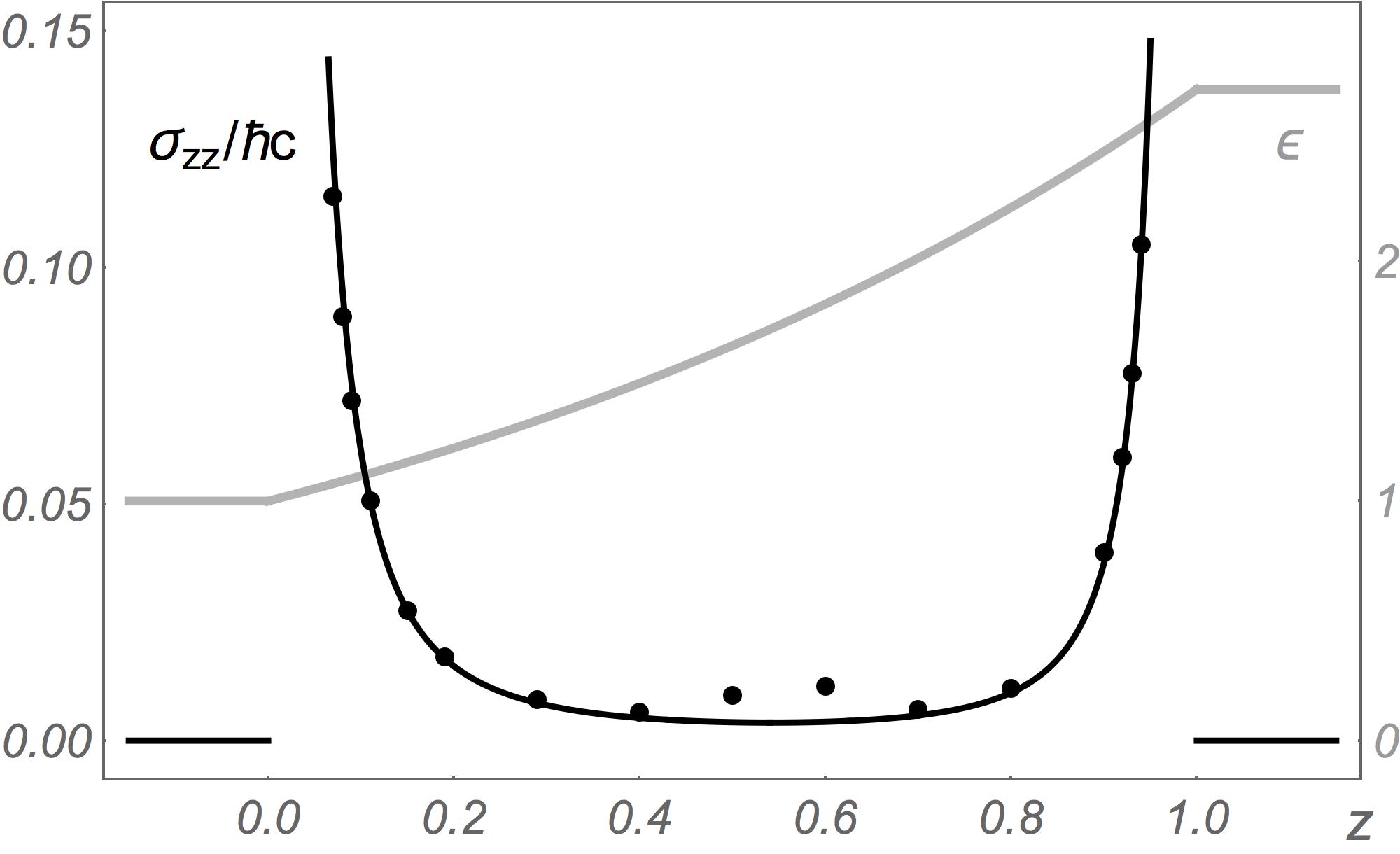}
\caption{
\small{
Casimir stress. Numerical computation (dots) of the Casimir stress $\sigma_{zz}$ for the profile $\varepsilon(z)$ of the electric permittivity shown (grey curve), $\mu=1$. The solid black curve shows the sum of our formula for the stress near each edge, Eq.~(\ref{final}), in excellent agreement with the numerical results near the edges. The stress is zero in the constant parts of the profile. We employed the profile \cite{PXL} $\varepsilon=\epsilon^z$ with $\epsilon=(\kappa^2+e\kappa_0^2)/(\kappa^2+\kappa_0^2)$ for $0<z<1$ and constant profiles outside. It includes Lorentzian-type dispersion for imaginary wavenumbers with real resonance at $\kappa_0=200$ (in the shown profile we put $\kappa=0$). Dispersion is necessary for the convergence of the Casimir stress \cite{Grin}.
}
\label{stressfigure}}
\end{center}
\end{figure}

{\it Theory.---} Consider a planar material that varies only in the $z$--direction. In this case, the Casimir--force density $\bm{\nabla}\cdot\sigma$ also points in the $z$--direction, while $\sigma$ is diagonal, such that $\sigma_{zz}$ is indeed the only physically relevant stress component. According to Lifshitz theory \cite{Grin,Lifshitz}: 
\begin{eqnarray}
\sigma_{zz} &=& -\frac{\hbar c}{(2\pi)^2} \int_0^\infty \int_0^\infty \left({\cal W}-{\cal W}_0\right)\,u\,du\,d\kappa \,,
\label{stress}\\
{\cal W} &=& \left. \sum_{\mathrm{p}=\mathrm{E},\mathrm{M}} \frac{1}{\nu_\mathrm{p}}\left(w^2-\partial_z\partial_{z_0}\right) \widetilde{g}_\mathrm{p}\, \right|_{z_0\rightarrow z} 
\label{fourier}
\end{eqnarray}
with $\kappa$ being the imaginary wave number and $u$ the spatial Fourier component. Going to imaginary wave numbers improves the convergence of the stress \cite{Review} as an integral of the spectral stress density ${\cal W}$ and, more importantly, describes the broadband nature of the Casimir effect, as each imaginary frequency requires a Hilbert transform of the material parameters over a wide range of real frequencies \cite{Review,Munday}. These parameters are the electric permittivity $\varepsilon$ and magnetic permeability $\mu$ that give rise to
\begin{equation}
n = \sqrt{\varepsilon\mu} \,,\quad \nu_\mathrm{E} = \mu \,,\quad \nu_\mathrm{M} = \varepsilon 
\end{equation}
for the two polarizations $\mathrm{E}$ and $\mathrm{M}$ of the electromagnetic field with Fourier--transformed Green functions $\widetilde{g}_\mathrm{p}$ satisfying the inhomogeneous wave equation
\begin{equation}
\partial_z \frac{1}{\nu_\mathrm{p}} \partial_z \widetilde{g}_\mathrm{p} - \frac{u^2+n^2\kappa^2}{\nu_\mathrm{p}} \widetilde{g}_\mathrm{p} = \delta(z-z_0) \,.
\label{Green}
\end{equation}
Note that ${\cal W}_0$ represents the diverging part in the spectral stress density of the electromagnetic zero--point fluctuations inside the material, which is removed in the renormalization of the Casimir stress, Eq.~(\ref{stress}). It is given by the geometrical--optics expression of the Green function that describes the outgoing waves from each point of the material \cite{Grin}. Figure \ref{stressfigure} shows the stress calculated numerically for a profile with discontinuities in the derivative of the refractive index. One clearly sees the divergencies at the edges. 

{\it Geometry.---} For getting an analytic expression of the characteristic behaviour we apply insights from geometry  --- transformation optics \cite{GREEBook} --- after two preluding remarks. First, the Casimir stress is given by the reflected waves inside the material and at its boundaries \cite{Grin}. Second, as we are employing waves with imaginary frequencies, their amplitudes are exponentially falling while propagating. Consequently, waves scattered from distant regions are exponentially suppressed. If there is a dominant contribution to the reflections, as due to the discontinuity of the derivative of the refractive index, we can thus replace the profile of the solid line Fig.~\ref{softwall} by one that also contains the characteristic feature --- the discontinuity of $dn/dz$ --- and does not cause scattering elsewhere. This is the profile of the Beltrami space \cite{Needham} (Fig.~\ref{softwall}, dotted line):
\begin{equation}
n = - \frac{b}{z} \quad\mbox{for} \quad z<0
\label{beltrami}
\end{equation}
that describes a maximally symmetric, open space \cite{Maximal} of constant negative curvature  for the electromagnetic field if, as in transformation optics \cite{GREEBook},
\begin{equation}
\varepsilon = \mu = n \,.
\label{geometry}
\end{equation}
Next we prove by direct calculation that this profile is scatteringless. Then we show that it remains so in the realistic case of $\varepsilon = n^2, \mu = 1$. To avoid clutter in our calculations we set the spatial units such that 
\begin{equation}
b = 1
\end{equation}
and reinstate units later. 

For the Beltrami profile of Eq.~(\ref{beltrami}) one can solve the equation for the Green function exactly:
\begin{equation}
g =  -\frac{e^{-\kappa s}}{2\pi\, c_+ c_-} \,,\quad
s = 2\, \mathrm{artanh}\,\frac{c_-}{c_+} 
\label{s} 
\end{equation}
where 
\begin{equation}
c_\pm = \sqrt{x^2+y^2+(z\pm z_0)^2} \,.
\end{equation}
One verifies that $g$ solves Eq.~(\ref{Green}) with $u^2=-\partial_x^2-\partial_y^2$ in physical space. One also verifies that $s$ satisfies the eikonal equation $(\bm{\nabla} s)^2 = n^2$, which proves that $s$ is the geodesic length --- the optical path length. From this follows that $g$ is exactly of the form required by geometrical optics, as it depends on frequency only through the exponential factor $\exp(-\kappa s)$ where $\kappa=-i\omega/c$ (with positive imaginary $\omega$ in our case): geometrical optics is exact for the Beltrami profile. Note that this is only true for the profile of Eq.~(\ref{beltrami}) from $-\infty$ to $0$ in its entirety. If $n$ turns to a constant at $z=-1$, forming an edge in the profile, the discontinuity in the refractive index will cause scattering, {\it i.e.}\ a violation of geometrical optics, and hence Casimir forces \cite{Grin}.

For the undisturbed profile of Eq.~(\ref{beltrami}) we solve for the Green function in Fourier space, Eq.~(\ref{Green}), and obtain for both polarizations:
\begin{equation}
\widetilde{g}= - I_\kappa(-uz_0) K_\kappa(-uz)
\label{geofourier}
\end{equation}
for $z<z_0$ and $z$ and $z_0$ interchanged for $z>z_0$, where $K$ and $I$ are the modified Bessel functions \cite{Erdelyi}. We will make use of this form in the case of realistic profiles with $\varepsilon\neq \mu$ where the interpretation of the material as establishing a geometry for the electromagnetic field is no longer exact \cite{GREEBook}. As it turns out, it will be still exact enough. 

{\it Reality.---} In the remainder of this paper we consider the realistic case of 
\begin{equation}
\varepsilon = n^2 \,,\quad \mu = 1 \,.
\label{realistic}
\end{equation}
In this case, the electric and magnetic properties of the material are different, and so the $\mathrm{E}$ and $\mathrm{M}$ polarizations differ as well:
\begin{equation}
\begin{split}
\widetilde{g}_\mathrm{E} = - \sqrt{z_0 z} \,I_\nu(-uz_0)K_\nu(-uz)\,, \\
\widetilde{g}_\mathrm{M} = - \frac{1}{\sqrt{z_0 z}} \,I_\nu(-uz_0)K_\nu(-uz)\,
\label{gEM}
\end{split}
\end{equation}
for $z<z_0$ and $z$ and $z_0$ interchanged for $z>z_0$, while we get for the index
\begin{equation}
\nu = \sqrt{\kappa^2 + 1/4} \,.
\label{index}
\end{equation}
The Green functions for the realistic case of Eq.~(\ref{realistic}) thus differ from Eq.~(\ref{geofourier}) of the geometric case of Eq.~(\ref{geometry}) by the prefactors $(z_0 z)^{1/2}$ and $(z_0 z)^{-1/2}$, respectively, which means that they are also scatteringless in space. However, as the index, Eq.~(\ref{index}), is different from $\kappa$, their temporal behaviour is modulated due to the different dependance on frequency $ic\kappa$: there is geometric dispersion \cite{Grin}. Yet for the renormalization of the Casimir stress in planar media, geometric dispersion is not relevant \cite{Grin}: we can thus regard the Green functions (\ref{gEM}) as describing the outgoing waves that give rise to ${\cal W}_0$ via Eq.~(\ref{fourier}) and are subtracted in the renormalization of the stress, Eq.~(\ref{stress}). 

Consider now the full profile of the soft wall (Fig.~\ref{softwall}) with $\varepsilon=\mu=1$ for $z<-1$ and the Beltrami profile of Eq.~(\ref{beltrami}) for $-1<z<0$. At the edge of the soft wall, $z=-1$, the derivative jumps from zero to $dn/dz=1$ in our units ($dn/dz=b^{-1}$ in general). For $-1<z_0<0$ and $z<z_0$ the Fourier--transformed Green functions are given by the outgoing waves of Eq.~(\ref{gEM}) plus a solution of the homogeneous wave equation:
\begin{equation}
\begin{split}
\widetilde{g}_\mathrm{E} = - \sqrt{z_0 z} \,I_\nu(-uz_0)\left[K_\nu(-uz) + \rho_\mathrm{E} I_\nu(-uz)\right],\\
\widetilde{g}_\mathrm{M} = - \frac{1}{\sqrt{z_0 z}} \,I_\nu(-uz_0)\left[K_\nu(-uz) + \rho_\mathrm{M} I_\nu(-uz)\right]
\label{ansatz}
\end{split}
\end{equation}
with coefficients $\rho_\mathrm{E}$ and  $\rho_\mathrm{M}$ for $-1<z$ and
\begin{equation}
\widetilde{g}_\mathrm{E} \propto e^{wz} \,,\quad \widetilde{g}_\mathrm{M} \propto e^{wz} \,,\quad w=\sqrt{\kappa^2+u^2}
\label{leftside}
\end{equation}
for $z<-1$. We see from Eq.~(\ref{Green}) that at $z=-1$ both $\widetilde{g}$ and $\partial_z\widetilde{g}$ must be continuous (the latter, because $\varepsilon$ and $\mu$ are continuous there). As the outgoing waves are the $I_\nu(-uz_0)K_\nu(-uz)$ waves we simply drop them in the renormalization and use only the reflected waves in Eqs.~(\ref{stress}) and (\ref{fourier}) of the Casimir stress. In this way we obtain
\begin{equation}
\sigma_{zz} = -\frac{\hbar c}{(2\pi)^2} \int_0^\infty \int_0^\infty \left(\rho_\mathrm{E}{\cal W}_\mathrm{E}+\rho_\mathrm{M}{\cal W}_\mathrm{M}\right) u\,du\,d\kappa
\label{szz}
\end{equation}
with ${\cal W}_\mathrm{E}$ and ${\cal W}_\mathrm{M}$ given by
\begin{equation}
\begin{split}
{\cal W}_\mathrm{E} = - \left(n^2\kappa^2+u^2-\partial_z\partial_{z_0}\right) \sqrt{z_0 z} \,H\bigg|_{z_0\rightarrow z} \,,\;\;\\
{\cal W}_\mathrm{M} = - \frac{1}{\varepsilon}\left(n^2\kappa^2+u^2-\partial_z\partial_{z_0}\right)\frac{1}{\sqrt{z_0 z}} \,H\bigg|_{z_0\rightarrow z}
\end{split}
\label{wEM}
\end{equation}
and $H=I_\nu(-uz_0)I_\nu(-uz)$. For evaluating the integrals in Eq.~(\ref{szz}) we use polar coordinates,
\begin{equation}
\kappa = w\cos\theta \,,\quad u = w\sin\theta \,,
\end{equation}
and the asymptotics of the integrand in the limit of $w\rightarrow\infty$, as a rapid growth of the stress in physical space corresponds to large components in Fourier space. We thus replace the modified Bessel functions by their asymptotics \cite{Erdelyi}:
\begin{equation}
\begin{split}
K_\nu(x) \sim \sqrt{\frac{\pi}{2}} \frac{e^{-\sqrt{\nu^2+x^2}+\nu\,\mathrm{arsinh} (\nu/x)}}{\sqrt[4]{\nu^2+x^2}} \,\\
I_\nu(x) \sim \frac{e^{\sqrt{\nu^2+x^2}-\nu\,\mathrm{arsinh} (\nu/x)}}{\sqrt{2\pi}\,\sqrt[4]{\nu^2+x^2}} \quad\quad\;\;
\label{asymptotics}
\end{split}
\end{equation}
and obtain for the ${\cal W}_\mathrm{E}$ and ${\cal W}_\mathrm{M}$ of Eq.~(\ref{wEM}) in the limit of $w\rightarrow\infty$ the expressions:
\begin{equation}
\begin{split}
{\cal W}_\mathrm{E} \sim - \frac{\cos^2\theta}{2\pi z\, [z^2-(z^2-1)\cos^2\theta]}\, e^{2w\phi(z)} \,,\\
{\cal W}_\mathrm{M} \sim \frac{2z^2+(1-2z^2)\cos^2\theta}{2\pi z\, [z^2-(z^2-1)\cos^2\theta]}\, e^{2w\phi(z)} \quad \,
\end{split}
\label{wEwM}
\end{equation}
with the exponent given by 
\begin{equation}
\phi(z) = \sqrt{\cos^2\theta+z^2\sin^2\theta} + \cos\theta\,\mathrm{arsinh}\frac{\cot\theta}{z} \,.
\label{phi}
\end{equation}
We also solve for $\rho_\mathrm{E}$ and $\rho_\mathrm{M}$ as follows: since $\partial_z\widetilde{g}=w\widetilde{g}$ for $\widetilde{g}$ of Eq.~(\ref{leftside}) for $z<-1$, continuity requires that the same is true for $\widetilde{g}$ of Eq.~(\ref{ansatz}) at $z=-1$, which establishes a linear equation for each $\rho_\mathrm{p}$. Using the asymptotics of the modified Bessel functions, Eq.~(\ref{asymptotics}), gives in the limit of $w\rightarrow \infty$:
\begin{equation}
\begin{split}
\rho_\mathrm{E} \sim - \frac{\pi\cos^2\theta}{4w}\, e^{-2w\phi(-1)} \,,\quad\;\;\\
\rho_\mathrm{M} \sim  \frac{\pi(2-\cos^2\theta)}{4w}\, e^{-2w\phi(-1)} \,.
\end{split}
\label{rho}
\end{equation}
Next we consider the asymptotics for $z\rightarrow -1$ in the integral of Eq.~(\ref{szz}) for the stress. The convergence of the integral is controlled by the exponents in Eqs.~(\ref{wEwM}) and (\ref{rho}), hence we take $\phi(z)-\phi(-1)\sim -(z+1)$ to first order in $z+1$ from Eq.~(\ref{phi}), while for the prefactors we put $z=-1$. We substitute $\zeta=\cos\theta$ and obtain
\begin{eqnarray}
\sigma_{zz} &=& \frac{\hbar c}{16\pi^2} \int_0^\infty e^{-2w(z+1)} w\,dw \int_0^1 (2-2\zeta^2 +\zeta^4) d\zeta \nonumber \\
&=& \frac{23}{960\pi^2}\, \frac{\hbar c}{(z+1)^2} \,.
\end{eqnarray}
Writing $a=z+1$ and reinstating units gives the main result of this paper, Eq.~(\ref{result}). It is elementary to generalise it to the case when a material with uniform refractive index $n_0$ different from unity meets a soft wall. We simply put the edge of the Beltrami profile $n=-1/z$ at $z=-1/n_0$ and express $n_0\kappa$ instead of $\kappa$ as $w\cos\theta$. We obtain along the same lines as above:
\begin{equation}
\sigma_{zz} = \frac{23n_0}{960\pi^2}\, \frac{\hbar c}{(z+n_0^{-1})^2} \,.
\end{equation}
As the first derivative of the Beltrami profile $-1/z$ is $n_0^2$ at $z=-1/z_0$, this corresponds to $b=1/n_0^2$. Hence we obtain in general units:
\begin{equation}
\sigma_{zz} = \frac{23}{240\,(2\pi)^2n_0^3}\, \frac{\hbar c}{a^2b^2} \,.
\label{final}
\end{equation}
Finally, in the case the first derivative of $n$ does not rise, but drops by $-b^{-1}$ at the edge we follow a similar procedure, and obtain the same result. 

Note that our result is only valid when dispersion, the frequency dependance of $\varepsilon$ and $\mu$, is not important in the relevant range of $w$. Ultimately, dispersion will soften the singularity of the Casimir stress near the edge, but it will not completely remove it, as the integral over the spatial Fourier components in Eq.~(\ref{stress}) remains divergent there. Our numerical results (Fig.~\ref{stressfigure}) show that our analytic formula, Eq.~(\ref{result}), describes well the intermediate regime near the edge until dispersion softens the power law. 

{\it Summary.---} For the first time, one can calculate the Casimir stress inside materials \cite{Grin}. We have found that the stress grows with a characteristic power law, Eq.~(\ref{result}), near the edge of a soft wall \cite{Soft} where the first derivative of the refractive index is discontinuous. The final formula, Eq.~(\ref{final}), represents one of the few analytic results in the theory of Casimir forces \cite{Analytic}. Our result also gives a first glimpse on new phenomena related to the aggregation of materials due to Casimir/ van der Waals forces at surfaces. Our paper answers the question of how such forces behave near edges of the refractive--index profile, but it also raises many more questions that may inspire future research. For example, what are stable configurations of aggregated materials? What are the time scales of aggregation? How does diffusion compete with Casimir/ van der Waals forces?

We thank 
Yehonathan Drori,
Mathias Fink,
and
Ephraim Shahmoon
for stimulating discussions. 
Itay Griniasty is grateful to the Azrieli Foundation for the award of an Azrieli Fellowship. Our work was also supported by ERC and ISF, a research grant from Mr.\ and Mrs.\ Louis Rosenmayer and from Mr.\ and Mrs.\ James Nathan, and the Murray B.\ Koffler Professorial Chair.



\begin{thebibliography}{99}
\bibitem{Remark}
The other non--vanishing components of the stress, $\sigma_{xx}$ and $\sigma_{yy}$, do diverge there, but they do not contribute to the force density $\bm{\nabla}\cdot\sigma$, as they depend only on $z$.
\end{thebibliography}
\end{document}